\documentclass[prl,aps,twocolumn,showpacs,bibnotes,superscriptaddress,epsf]{revtex4}%superscriptaddress,

\usepackage{float}
\usepackage{graphics}
\usepackage{graphicx}% Include figure files
\usepackage{dcolumn}% Align table columns on decimal point
\usepackage{bm}% bold math
\usepackage{SIunits}
\usepackage{tabularx}
\usepackage{subfigure}
\usepackage{tabularx}
\usepackage{amsmath}
\usepackage[urlcolor=blue]{hyperref}
\hypersetup{backref, colorlinks=true, linkcolor=blue, citecolor=blue}

\begin{document}

\title{Nontrivial Metallic State of Molybdenum Disulfide}

\author{Zi-Yu Cao}
\affiliation{Key Laboratory of Materials Physics, Institute of Solid State Physics, Chinese Academy of Sciences, Hefei 230031, China}
\affiliation{University of Science and Technology of China, Hefei 230026, China}
\affiliation{Center for High Pressure Science and Technology Advanced Research, Shanghai 201203, China}

\author{Jia-Wei Hu}
\affiliation{Center for High Pressure Science and Technology Advanced Research, Shanghai 201203, China}

\author{Alexander F. Goncharov}
\affiliation{Key Laboratory of Materials Physics, Institute of Solid State Physics, Chinese Academy of Sciences, Hefei 230031, China}
\affiliation{University of Science and Technology of China, Hefei 230026, China}
\affiliation{Geophysical Laboratory, Carnegie Institution of Washington, Washington, DC 20015, U.S.A.}

\author{Xiao-Jia Chen}
\email{xjchen@hpstar.ac.cn}
\affiliation{Center for High Pressure Science and Technology Advanced Research, Shanghai 201203, China}

\date{\today}

\begin{abstract}
The electrical conductivity and Raman spectroscopy measurements have been performed on MoS$_2$ at high pressures up to 90 GPa and variable temperatures down to 5 K. We find that the temperature dependence of the resistance in a metallic 2\emph{H}$_a$ phase has an anomaly (a hump) which shifts with pressure to higher temperature. Concomitantly, a new Raman phonon mode appears in the metallic state suggesting that the electrical resistance anomaly may be related to a structural transformation. We suggest that this anomalous behavior is due to a charge density wave state, the presence of which is indicative for a possibility for an emergence of superconductivity at higher pressures.
\end{abstract}


\maketitle

Two-dimensional (2D) materials are characterized by a strong anisotropy of the intra- versus inter- layer bonding. This remarkable difference brings up unique notions such as 2D materials which can be exploited for practical applications \cite{Novo}. One of the most exposed materials of such a kind is graphene \cite{Novo,Novos}, which also brought the interests to inorganic materials with unique electronic \cite{Radi}, optical \cite{Wang} attributes and high mobility \cite{bolo}. However, graphene is a gapless (semimetallic) material which restricts its possible applications. In contrast, the transition-metal dichalcogenides (TMDs), quasi-2D material with distinct structures and unique physical properties \cite{Fung,Guil,Revo,Ali,Qian}, possess a non-zero band gap that could also be tuned. Therefore, TMDs are currently attracted enormous research interests. MoS$_2$ is one of the most extensively investigated member of TMDs, and it can be prepared as a 2D material via mechanical exfoliation \cite{Wang,Mak}. By varying the number of layers, MoS$_2$ can be transformed from an indirect band-gap semiconductor to a direct band-gap semiconductor \cite{Mak}. Monolayer MoS$_2$ has mechanical properties almost as good as graphene, but unlike graphene, possesses a direct bandgap. Thus, MoS$_2$ shows an excellent potential to replace graphene in the next generation nanoelectronics applications \cite{kfMak,Zeng,Xiao,Xiao-2,Zhu}. Furthermore, doping of 2D MoS$_2$ has been shown to change the electronic and structural properties dramatically that results to metallization, formation of the charge density wave (CDW) \cite{Zhuang} and eventually superconducting states \cite{JTYe,Yuan}. Similar structural effects related to 2\emph{H} to 1\emph{T}¡¯ transition (from trigonal to octahedral TM coordination) have been found in other doped TMDs (MoTe$_2$), confirmed by Raman spectroscopy measurements \cite{Wang-2} including the case when 1\emph{T}¡¯ phase has been stabilized by pressure and superconductivity appears accordingly \cite{Qi}.

The application of external pressure is an alternative (to doping) way to tune the electronic as well as crystal structure. MoS$_2$ transforms from 2\emph{H}$_c$ to 2\emph{H}$_a$ phase (through a sliding of the layers) above 20 GPa along with a metallization \cite{Chi,Hromadov¨¢}. However, the resistivity measurements show that the temperature dependence of the resistivity is not monotonous in both semiconducting and metallic regimes revealing a pressure dependent hump in the resistivity$-$temperature curves \cite{Chi}. This suggests that an additional electronic or structural ordering in MoS$_2$ at low temperatures and high pressures. It is common for TMDs with both major structural types 2\emph{H} and 1\emph{T} to demonstrate CDW phenomena, which lower the symmetry of their quasi-2D structures \cite{Sugai,Holy,Sugai1,Sugai2,Scott}. It has been noticed that the presence of CDW indicates the proximity of superconducting state emerging at lower temperatures. These two states have been shown to demonstrate the competing orders which could also be applied to high-temperature superconductors \cite{Wu}. The highest superconducting critical temperature is observed in materials with CDW ordering that occurs at the comparable temperatures \cite{Wilson,Castro}. Superconductivity in MoS$_2$ has been observed for heavily doped 2D materials \cite{JTYe} but no CDW state has not been reported. It remains largely unclear whether superconductivity in 2\emph{H}-MoS$_2$ could be induced by application of high pressure \cite{Chi-2} and will it be related to CDW state.

In this work, we report on the electrical transport and Raman spectroscopy measurements of MoS$_2$ at high pressures and low temperatures. Anomalies are discovered in both the resistance$-$temperature (R-T) curves and the Raman spectra. A hump in the R-T dependence has been observed in both semiconducting and metallic states increasing in temperature with pressure. In the metallic phase a new Raman band appears increasing in intensity with pressure and then diminishing and on cooling down. We connect the R-T hump and this new Raman feature as well as an anomalous behavior of the regular phonon modes with developing of a CDW order on MoS$_2$ at high pressure by analogy with observations of similar states in other (commonly heavier) TMDs. Furthermore, this order is indicative for a possible superconducting order at higher pressures.

Four MoS$_2$ samples have been investigated in separate electrical resistance (A) and Raman (B, C, and D) experiments. For electrical resistance (A) measurements, a miniature diamond anvil cell with anvils in 300 \emph{$\mu$m} culet was used. A sample chamber of diameter 120 \emph{$\mu$m} was formed by drilling in the c-BN gasket positioned then situated between tips of two diamond anvils. Four Pt wires were adhered to the sample with silver epoxy by the Van Der Pauw method \cite{Van} to measure the in plane resistance. Daphne oil 7373 was used as a pressure transmitting medium to ensure hydrostatic pressure conditions in the sample chamber. The pressure was determined by the spectral shift of the ruby fluorescence R1 peak. Resistance was measured in the Physical Property Measurement System (by Quantum Design). For Raman (B, C, and D) experiments, a Mao-Bell diamond anvil cell with 200 \emph{$\mu$m} culet, low fluorescence diamonds, and rhenium gasket combined with a He continuous flow cryostat was used for high-pressure and low-temperature Raman measurements. Neon was loaded as the pressure transmitting medium. Pressure was measured in situ at high pressure and low temperatures using a ruby fluorescence technique. The temperature was measured using a Pt resistance sensors attached to the diamond anvil cell close to the sample with a typical precision of $\pm$0.5 K. The Raman spectra were measured using a single stage spectrograph equipped with an array thermoelectrically cooled CCD detector. The Raman notch filters were of a very narrow bandpass (Optigrate) allowing Raman measurements down to 10 cm$^{-1}$ in the Stokes and AntiStokes \cite{Gonc}. The 488 nm excitation line was used to illuminate an approximately 5 \emph{$\mu$m} x 5 \emph{$\mu$m} spot on the newly cleaved shining surface of MoS$_2$. The Raman spectra were measured with laser power down to 1 mW to avoid laser heating effects.

\begin{figure}[tbp]
\includegraphics[width=\columnwidth]{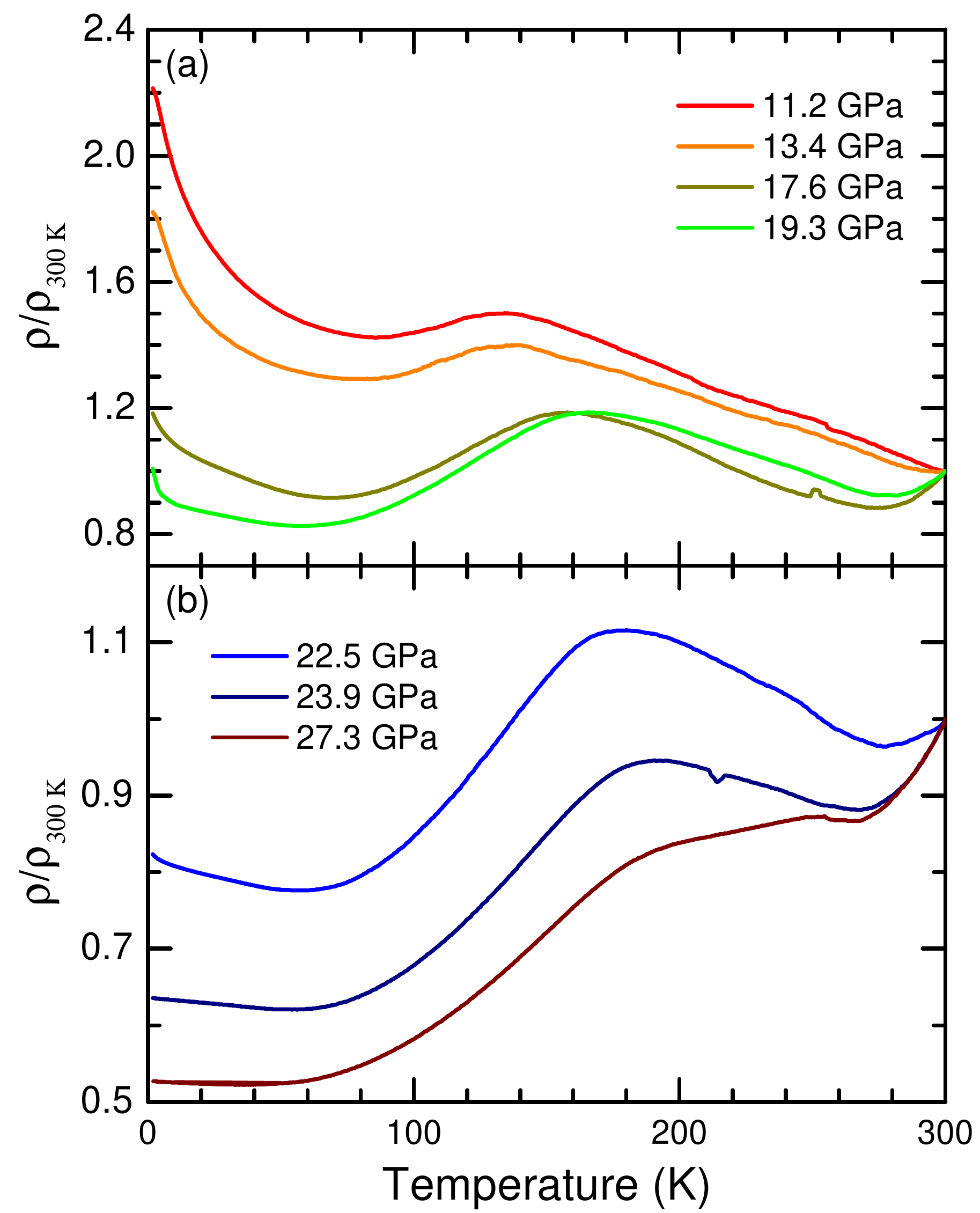}
\caption{Temperature dependence of the resistance of MoS$_2$ at various pressures measured at 300 K. (a) Semiconducting behavior at pressures below 20 GPa. (b) Metal behavior at pressures between 20 GPa and 30 GPa.}
\end{figure}

Figures 1(a) and 1(b) display the R-T curves obtained at various pressures around the semiconductor-metal transition. The resistance declines by 3 magnitude with increasing pressure up to 27.3 GPa at room temperature. Below 20 GPa, the R-T curves exhibit an overall semiconducting behavior, but a hump is observed reversing the slope in the temperature interval below the maximum. When pressure is increased above 20 GPa, a phase transition from 2\emph{H}$_c$ to 2\emph{H}$_a$ occurs \cite{Chi,Hromadov¨¢}, and the system turns into a metallic state \cite{Nayak,Shen}. However, we find that above 20 GPa, the R-T curves show an anomaly, where a generally metallic behavior turns into a semiconducting one at temperatures above the hump. Moreover, the R-T slope is nearly zero (if not negative) in the limit of low temperatures. The hump is moving to higher temperatures with pressure.

\begin{figure*}[tbp]
\includegraphics[width=2\columnwidth]{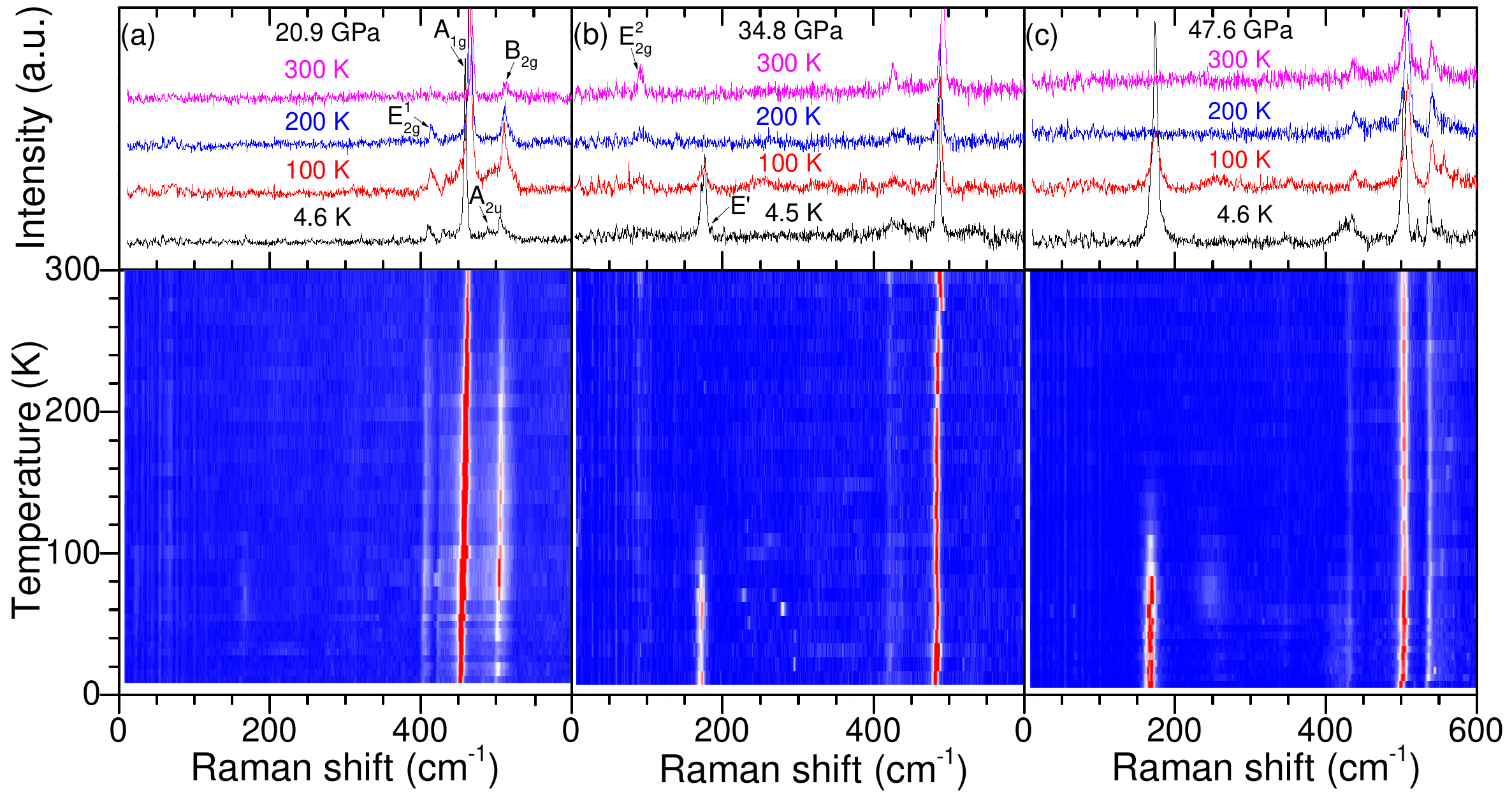}
\caption{Raman data of MoS$_2$ at various temperatures at 20.9 GPa (sample B) (a), 34.8 GPa (sample C) (b), and 47.6 GPa (sample B) (c). The upper panels are the spectra at various temperatures and the bottom panel is the 2D presentation in the Raman frequency$-$temperature coordinates.}
\end{figure*}

The presence of a hump in the R-T curve is a common behavior for TMDs and it was related to the presence of CDW state \cite{Naito}, which results in a crossover on the resistance behavior due to the electronic scattering by CDW fluctuation changing the R-T slope and character at low temperatures. This behavior can also be considered as due to the fact that a portion of the Fermi surface becomes gapped in CDW state \cite{Peierls}. As a quasi-2D compound, MoS$_2$ has immense potential to enter into CDW state below certain temperature. Thus, CDW state is a possible candidate in the observed nontrivial metallic state (Fig. 1).

\begin{figure*}[t]
\includegraphics[width=2\columnwidth]{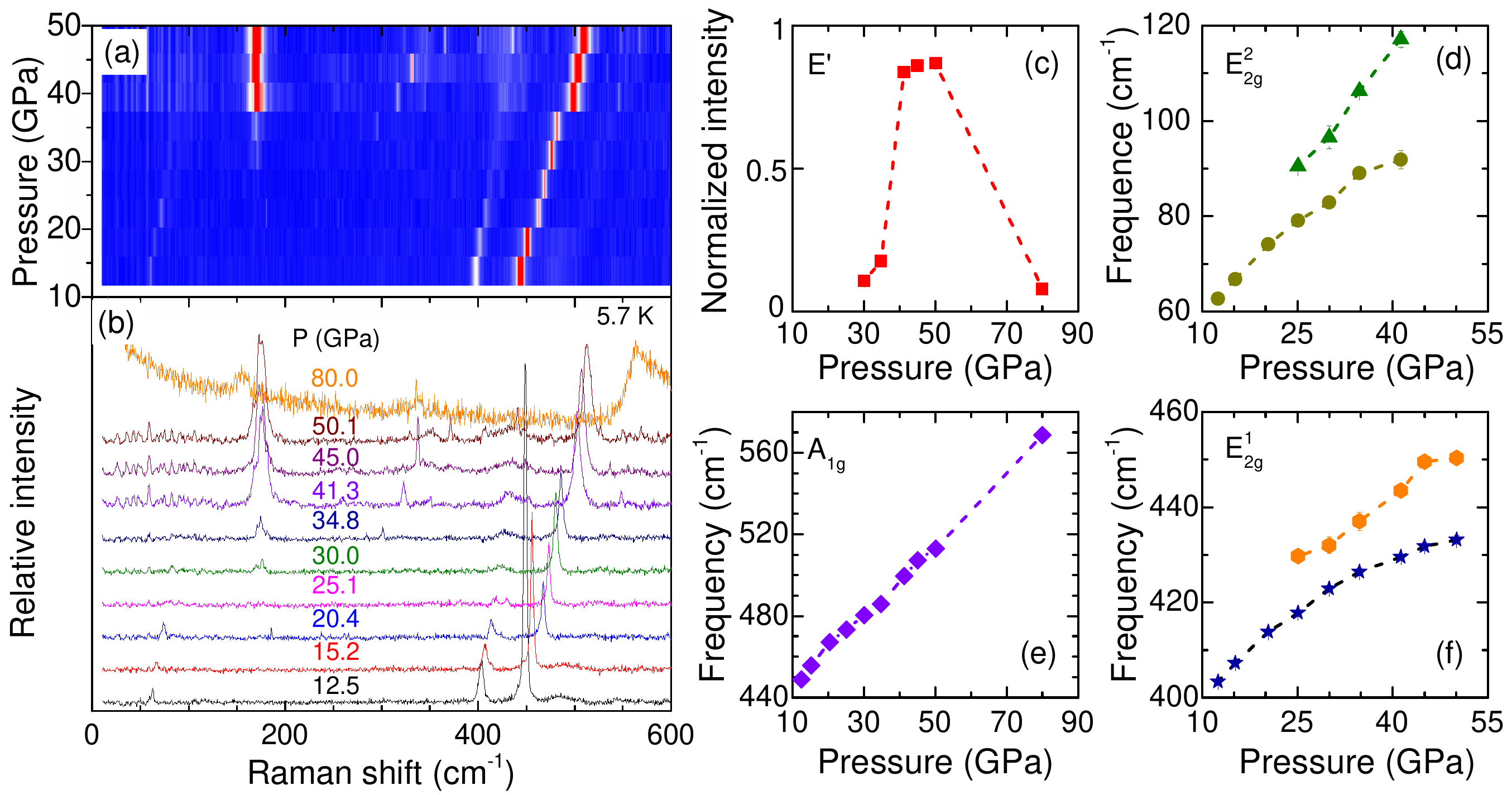}
\caption{Raman data of MoS$_2$ at temperature of 5.7 K and various pressures up to 80.0 GPa. (a) The 2D map of the Raman intensity at 5.7 K in the frequency-temperature coordinates. (b) Raman spectra at various pressures at 5.7 K. (c) The pressure dependence of the intensity of the E$'$ mode at 5.7 K. The pressure dependence of the Raman frequencies of the E$_{2g}^2$ (d), A$_{1g}$ (e), and E$_{2g}^1$ (f) modes at 5.7 K.}
\end{figure*}

For better understanding of the nontrivial metallic state, we performed Raman measurements in the pressure-temperature range of interest near the 2\emph{H}$_\emph{c}$ to 2\emph{H}$_\emph{a}$ transition and in high-pressure phase up to 90 GPa (Figs. 2-4). The 2\emph{H}$_\emph{c}$ to 2\emph{H}$_\emph{a}$ transition is isostructural, so the same Raman fundamental modes are expected to be observed in the spectra: A$_{1g}$+E$_{1g}$+2E$_{2g}$. At ambient conditions these modes are observed at 32 (E$_{2g}^2$), 287 (E$_{1g}$), 381 (E$_{2g}^1$), and 409 (A$_{1g}$) cm$^{-1}$ \cite{Wilson-3}. Our low temperature Raman spectra demonstrate that these modes behave quite similar to the observations reported at room temperature \cite{Chi}. However, the E$_{1g}$ and E$^2_{2g}$ modes are weak (see also Refs. \cite{Sugai,Sugai2} for other TMDs), and could not be observed in some of our spectra. In addition, in some Raman experiments (e.g. sample B) we have observed Raman-forbidden modes above the position of the A$_{1g}$ mode, which we tentatively assigned to out of plane A$_{2u}$ and B$_{2g}$ infrared and inactive modes \cite{Molina}, respectively. As in Ref. \cite{Chi}, we find that E$_{2g}^2$ and E$_{2g}^1$ modes show discontinuities in frequency, while the A$_{1g}$ mode changes the pressure slope demonstrating the 2\emph{H}$_\emph{c}$ to 2\emph{H}$_\emph{a}$ transition. At 5.7 K the transition starts at 25 GPa and is completed at nearly 50 GPa compared to the 20-40 GPa range at room temperature \cite{Chi}. Based on the results of previous experiments \cite{Chi} our temperature runs at 20.9, 34.8, and 47.6 GPa (Fig. 2) should be referred to semiconducting 2\emph{H}$_\emph{c}$, mixed 2\emph{H}$_\emph{c}$ and 2\emph{H}$_\emph{a}$, and mainly metallic 2\emph{H}$_\emph{a}$ phases, respectively.

\begin{figure}[bp]
\includegraphics[width=\columnwidth]{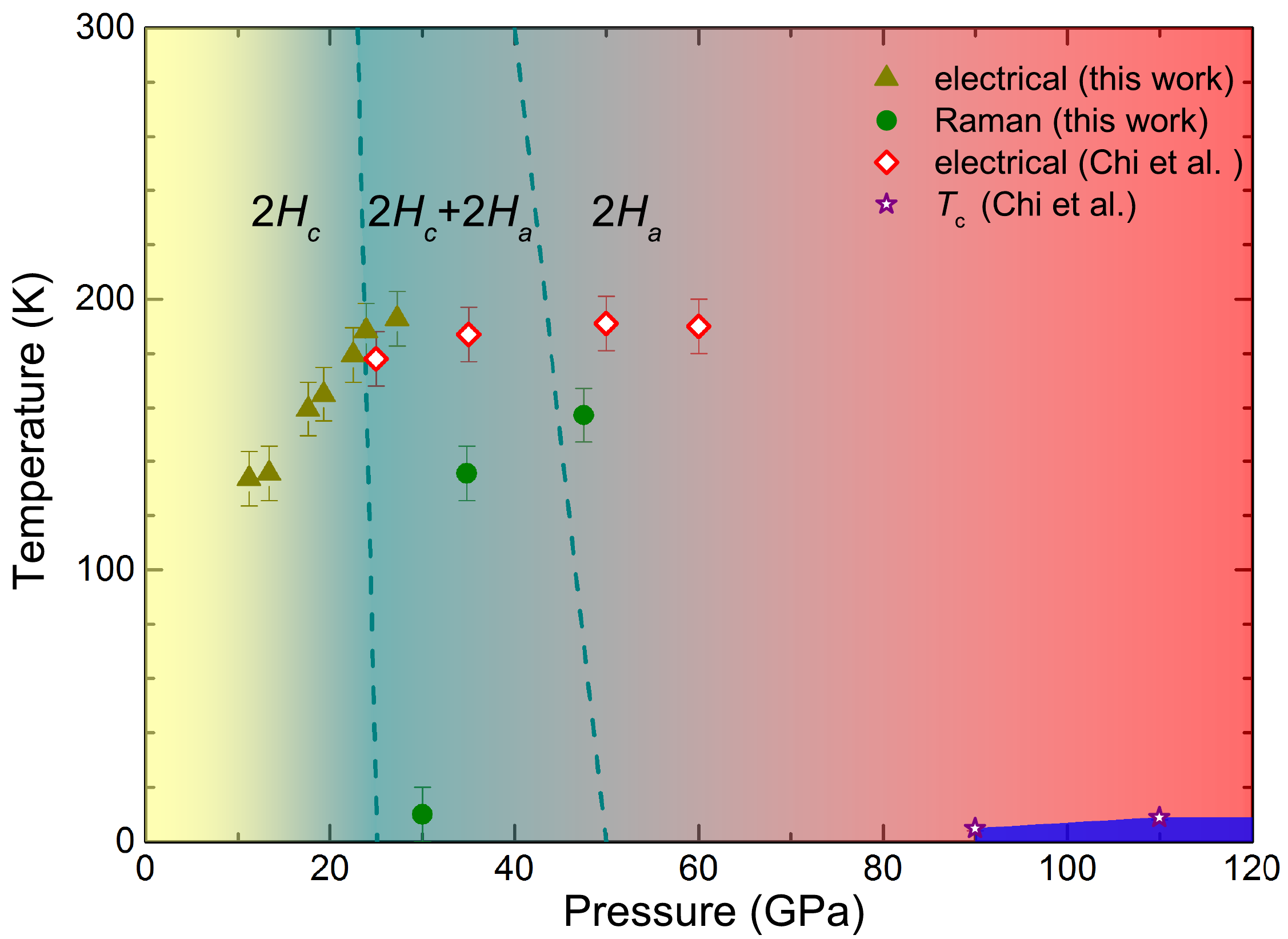}
\caption{Phase diagram illustrating the results, see legends. The results of this work are the filled symbols. The results from the works of Chi {\it et al}. \cite{Chi,Chi-2} are open symbols. The dark cyan dot-dashed lines are the phase lines (separating pure 2\emph{H}$_c$, mixed, and 2\emph{H}$_a$ phases determined from Raman spectroscopy in this work at 5.7 K and Ref. \cite{Chi} at 297 K).}
\end{figure}

A new mode at 174 cm$^{-1}$ (E$'$) appears in the metallic 2\emph{H}$_\emph{a}$ phase at 30 GPa at 5.7 K and increases in intensity with pressure. This mode becomes very strong at 40$-$50 GPa, but drops in intensity at higher pressures. The frequency of this mode is almost pressure independent up to 50 GPa, but it softens at higher pressures reaching 156 cm$^{-1}$ at 80 GPa$-$the highest pressure investigated here. In the temperature scans at 35$-$50 GPa, the intensity of this E$'$ mode stays almost constant below approximately 60 K, and then drops down so the mode nearly vanishes above 130$-$150 K (Fig. 3). The frequency of the E$'$ mode remains essentially unchanged. The origin of this new E$'$ peak is likely a phonon mode as will be discussed below. Concomitantly, other phonon modes show anomalous temperature dependencies. The A$_{1g}$ mode shows a softening, while the E$^1_{2g}$ modes dramatically broadens. These anomalous changes suggest the presence of an additional phase transition or electronic ordering, which we tentatively assign to CDW. The lock in temperature related to this phonon anomaly, \emph{T}$_{\emph{CDW}}$, is comparable to that where the hump in electrical resistance appears at pressures above 50 GPa (Fig. 4), but disagrees at lower pressures. This is likely because the phonon anomalies can only be observed in the metallic 2\emph{H}$_\emph{a}$ phase states (including mixed 2\emph{H}$_\emph{c}$-2\emph{H}$_\emph{a}$), while the conductivity hump could be observed in the 2\emph{H}$_\emph{c}$ semiconducting and 2\emph{H}$_\emph{c}$-2\emph{H}$_\emph{a}$ mixed phase samples.

A newly observed E$'$ mode is in a good correspondence in frequency with a transverse acoustic modes near the Brillouin zone boundaries at the K and M points \cite{Molina,Waka}. The position of this mode at ambient pressure is near 170 cm$^{-1}$. However, it is difficult to judge about the position of the zone boundary acoustic modes because their pressure dependence is unknown in the semiconducting 2\emph{H}$_\emph{c}$ phase and there may be a discontinuity at the layer sliding 2\emph{H}$_\emph{c}$-2\emph{H}$_\emph{a}$ transition. Assuming that the E$'$ mode has the phonon origin and given its weak temperature dependence one can propose that the appearance of this mode in the Raman spectra manifests the development of a superstructure, which results in the Brillouin zone folding \cite{Holy}. Apart of an obvious possibility of a structural phase transition, CDW has been shown to produce periodic lattice distortion leading to observations of nominally forbidden Raman bands.

Theoretical researches have discussed the possibility of CDW in MoS$_2$ \cite{Zhuang}, but in relations to another underlying crystal structure-1\emph{T}, which makes the results not immediately applicable to high-pressure 2\emph{H}$_\emph{a}$ polymorph investigated here. The closest in properties TMDs are 2\emph{H}-NbS$_2$ and NbSe$_2$, and both reveal interesting phenomena at low temperatures \cite{Mas}, including CDW and superconductivity. Raman spectra of 2\emph{H}-NbSe$_2$ through the transformation to CDW and have been extensively studied \cite{Mas,Tsang,Mialitsin}. They demonstrate strongly temperature dependent excitations (the amplitude modes) and also essentially temperature independent ¡°frozen¡± in modes, the nature of which is often unclear.

CDW and superconductivity in TMDs normally coexist and often are considered to be the competing orders. Thus, one can expect superconducting state to develop in the proximity of CDW state. It is interesting that 2\emph{H}-NbS$_2$ was considered as an exception as in this material CDW is not observed while superconductivity has been found at 6 K. This justifies to importance of studying competing CDW and superconducting \cite{Chi-2} orders at high pressures.

In conclusion, we have reported on electrical transport, and Raman scattering measurements of MoS$_2$ at high pressures and low temperatures. The temperature dependencies of the electrical conductivity show anomalies (hump) in semiconducting and metallic states. The Raman spectra also show anomalous behavior in the metallic state revealing a new peak at low temperatures along with the phonon mode softening and broadening. These behaviors suggest the presence of an additional order at low temperature, which creates a modification of the electronic structure (gapping of the Fermi surface) and structural instabilities (CDW). Given a common coexistence and competing of CDW and superconductivity in TMDs, an unusual superconductivity at very high pressures ($>$80 GPa) is justified.

\end{document}